\documentstyle[multicol,aps,prb,epsf]{revtex} 

\begin{document}
\author{G. Gomila}
\address{Centre de Recerca en Bioelectr\`{o}nica i Nanobioci\`{e}ncia and
Departamentd'Electr\`{o}nica-Universitat de Barcelona, Laboratori de
Nanobioenginyeria-Parc Cient\'{\i}fic de Barcelona, C/ Josep Samitier 1-5,
E-08028 Barcelona, Spain. }
\author{I. R. Cantalapiedra}
\address{Departament de F\'{i}sica Aplicada, Universitat Polit\`{e}cnica de
Catalunya, Av. Dr. Mara\~{n}on 44, E-08028 Barcelona, Spain. }
\author{L. Reggiani}
\address{INFM-National Nanotechnology Laboratory and Dipartimento di Ingegneria
dell'Innovazione, Universit\`{a} di Lecce, Via Arnesano s/n, I-73100,
Lecce,Italy.}
\title{Unipolar transport and shot noise in metal-semiconductor-metal structures}
\maketitle

\begin{abstract}
We carry out a self-consistent analytical theory of unipolar current and
noise properties of metal-semiconductor-metal structures made of highly
resistive semiconductors in the presence of an applied bias of arbitrary
strength. By including the effects of the diffusion current we succeed to
study the whole range of carrier injection conditions going from low level
injection, where the structure behaves as a linear resistor, to high level
injection, where the structure behaves as a space charge limited diode. We
show that these structures display shot noise at the highest voltages.
Remarkably the crossover from Nyquist noise to shot noise exhibits a
complicate behavior with increasing current where an initial square root
dependence (double thermal noise) is followed by a cubic power law.
\end{abstract}


\smallskip 
\begin{multicols}{2}

\section{Introduction}

Metal-semiconductor-metal structures have been of fundamental and applied
interest since the birth of the physics of semiconductor device. Here we
concentrate on the relevant case of high resistivity semiconductors where
metals usually form Ohmic or injecting contacts.\cite{Sze81} A common
feature of these structures is that carrier transport is mainly limited by
diffusion, with thermionic emission playing a negligible role. Furthermore,
when a bias voltage is applied to these structures, they can behave either
as a linear resistor (if carrier injection from the contacts is negligibly
small) or as non-linear space charge limited device (if carrier injection
from the contacts is extremely high).\cite{Lampert72} Therefore, these
structures are still offering excellent opportunities to study current noise
properties induced by diffusion noise in a variety of transport conditions.

The present knowledge of the noise properties of these structures can be
summarized as follows. At thermodynamic equilibrium, the low frequency
spectral density of current fluctuations, $S_{I}^{eq}(0)$, is given by 
\begin{equation}
S_{I}^{eq}(0)=\frac{4k_{B}T}{R^{eq}},  \label{Nyquisteq}
\end{equation}
in agreement with Nyquist theorem.\cite{Nyquist28} Here, $R^{eq}$ is the
equilibrium device resistance, $T$ the bath temperature and $k_{B}$ the
Boltzmann constant.

Out of equilibrium, when an external bias (voltage or current) is applied,
Nyquist relation no longer holds and deviations from Eq.(\ref{Nyquisteq})
are expected. These deviations have been studied in detail in two limiting
cases, namely: (i) when the structure behaves as a space charge limited
diode (with a strong inhomogeneous profile of the free carrier density), 
\cite{Vliet75,Nicolet75,Ziel79} and (ii) when the structure behaves as a
linear resistor (with a homogeneous profile of the free carrier density).%
\cite{Gomila00b}

In the former case (i) the current $I$ displays a quadratic dependence on
the applied voltage $V$ (Mott-Gurney law) $I=\beta V^{2}$, where $\beta $ is
a sample dependent parameter. Accordingly, the low frequency spectral
density of current fluctuations was found to take the form\cite
{Vliet75,Nicolet75,Ziel79} 
\begin{equation}
S_{I}(0)=\frac{8k_{B}T}{Z(0)}=16k_{B}T\beta ^{1/2}I^{1/2},  \label{Isquare}
\end{equation}
where $Z(0)$ is the low frequency impedance (differential resistance). From
the form of the first equality, this type of noise was called double thermal
noise.\cite{Nougier78} The cross-over between Nyquist noise and high voltage
space charge limited conditions is summarized by the formula 
\begin{equation}
S_{I}(0)=4k_{B}T\frac{V}{I}\left( \frac{dI}{dV}\right) ^{2}.  \label{Double}
\end{equation}
Highly resistive structures under strong carrier injection conditions have
been used to successfully test experimentally this prediction.\cite
{Nougier78}

In the latter case (ii) the structure behaves as a linear resistor and
displays a resistance $R=L/(qA\mu \overline{n})$, where $A$ is the cross
sectional area, $q$ the carrier charge, $\mu$ the mobility and $\overline{n}$
the free carrier density (the bar denotes an average with respect to
fluctuations). When the applied bias is high enough as to make the transit
time due to drift $\tau _{T}=L^{2}/\mu V$ significantly shorter than the
dielectric relaxation time $\tau _{d}=\epsilon /q\mu \overline{n}$, where $%
\epsilon$ is the dielectric permittivity, the structure was predicted to
exhibit shot noise \cite{Gomila00b} with 
\begin{equation}
S_{I}(0)=2q\overline{I} .  \label{shotI}
\end{equation}

The above results show two completely different behaviors for the
non-equilibrium current noise properties of metal-semiconductor-metal
structures depending on whether the stationary free carrier distribution is
strongly inhomogeneous (double thermal noise) or homogeneous (shot noise).
The physical reason for such a discrepancy of behavior is an intriguing
problem at present. In the authors opinion its origin can be traced back to
the theoretical difficulty of accounting for the effects of the diffusion
current on the non-equilibrium current fluctuations in the whole range of
different carrier injection conditions.

The purpose of the present paper is to overcome such a difficulty by
developing an analytical theory for the non-equilibrium current noise
properties of metal-semiconductor-metal structures made of highly resistive
semiconductors that includes properly the effects of the diffusion current.
The theory is valid in the whole range of physical conditions going from
homogeneous to highly inhomogeneous free carrier distributions, thus
allowing to describe in a continuous manner the current spectral density as
a function of the applied bias for different levels of current injection.

The paper is organized as follows. Section \ref{System} describes the system
under study. Section \ref{Model} presents the physical model used to
describe the low frequency transport and the non-equilibrium current
fluctuation properties. Section \ref{Solution} is devoted to analyze the
charge transport properties of the device, while Section \ref{Results_Noise}
analyzes its non-equilibrium current noise. In particular, we analyze the
effects of space charge on the shot noise properties. Section \ref
{Conclusions} draws the main conclusions of our investigation. Two
appendices provide the most technical derivations.

\section{System under study}

\label{System} The system under study consists of an active region made by a
non-degenerate $n$-type semiconducting material sandwiched between two metal
plates that act as contacts. The semiconductor is assumed to be lightly
doped with a donor density $N_{D}$. All donors are assumed to be ionized at
the considered temperatures (the case of a $p$-type semiconductor can be
analogously considered). To justify a one-dimensional electrostatic
treatment in the $x$ direction and to neglect the effects of the boundaries
in the $y$ and $z$ directions the transversal size of the sample is assumed
to be much larger than the characteristic Debye screening length. The metals
are assumed to form ohmic injecting contacts so that the voltage drop inside
them, or equivalently, the contact resistance, can be neglected to a good
approximation. Accordingly, when a voltage is applied to the structure, all
the potential drop takes place inside the semiconductor between positions $%
x=0$ and $x=L$, and the contacts can be excluded from consideration.
Depending on the parameter values of the contacts homogeneous as well as
inhomogeneous conditions will be studied.

We assume that inelastic scattering processes with phonons are dominant, so
that carrier thermalization at the bath temperature holds at any point.
Accordingly, for the applied bias considered here no carrier heating takes
place which allows us to use a field independent electron mobility and
diffusivity.

\section{Physical model}

\label{Model}

\subsection{Charge transport}

\label{Model_Transport}

The transport approach appropriate to describe the electrical properties of
the metal-semiconductor-metal structure under study consists of the standard
drift-diffusion current equation self-consistently coupled to the Poisson
equation and supplemented by appropriate boundary conditions. The current
drift-diffusion equation reads\cite{Sze81} 
\begin{equation}
\frac{\overline{I}}{A}=q\overline{n}(x)\mu \overline{E}(x)+qD\frac{d%
\overline{n}(x)}{dx}\text{,}  \label{DriftDif}
\end{equation}
where $A$ is the cross sectional area, $\overline{n}(x)$ the local carrier
density, $\mu $ the mobility (assumed to be field independent), $\overline{E}%
(x)$ the local electric field and $D$ the diffusion coefficient (related to
the mobility through Einstein's relation $D/\mu =k_{B}T/q$, where
non-degenerate statistics is assumed). The bar denotes an average with
respect to fluctuations. The Poisson equation is as usual 
\begin{equation}
\frac{d\overline{E}(x)}{dx}=\frac{q}{\epsilon }\left[ N_{D}-\overline{n}%
(x)\right] ,  \label{Poisson}
\end{equation}
with $\epsilon $ the static dielectric constant of the semiconducting
material. The appropriate boundary conditions to describe the ohmic
injecting contacts are given by (see appendix \ref{metal-semiconductor}) 
\begin{equation}
\overline{n}(0)=\overline{n}(L)=n_{c},  \label{Bc}
\end{equation}
where the carrier density at the contacts $n_{c}$ is independent of the
applied bias and given by (see appendix \ref{metal-semiconductor}) 
\begin{equation}
n_{c}=N_{C}\exp \left( -\frac{q\phi _{bn}}{k_{B}T}\right) .  \label{nc}
\end{equation}
Here, $N_{C}$ is the effective density of states in the conduction band and $%
\phi _{bn}$ the metal-semiconductor barrier height. For values of the
contact parameters such as $n_{c}=N_{D}$ the stationary free carrier
distribution is homogeneous, while for values such as $n_{c}>N_{D}$ there is
net carrier injection from the contacts and the stationary carrier
distribution is inhomogeneous.

Equations (\ref{DriftDif}) and (\ref{Poisson}) can be combined into a single
equation for the electric field 
\begin{equation}
-D\frac{d^{2}\overline{E}(x)}{dx^{2}}-\mu \overline{E}(x)\left[ \frac{d%
\overline{E}(x)}{dx}-\frac{qN_{D}}{\epsilon }\right] =\frac{\overline{I}}{A}
\label{DDE_x}
\end{equation}
subject to the boundary conditions 
\begin{equation}
\frac{d\overline{E}(0)}{dx}= \frac{d\overline{E}(L)}{dx}=\frac{q}{\epsilon }%
\left[ N_{D}-n_{c}\right].  \label{BcE_x}
\end{equation}

\subsection{Current fluctuations}

\label{Model_Fluctuations} The only source of fluctuations in the system is
related to the diffusion of carriers inside the structure. Accordingly, the
low frequency noise properties (beyond $1/f$ noise) can be described through
a drift-diffusion-Langevin model.\cite{Ziel74,Vliet94,Bonani01} It consists
of the linearized version of the transport model presented in the previous
subsection supplemented by the appropriate Langevin source which describes
the diffusion noise. The current equation in explicit form reads, \cite
{Ziel74,Vliet94,Bonani01} 
\begin{eqnarray}
\frac{\delta I(t)}{A} &=&q\mu \overline{E}(x)\delta n_{x}(t)+q\overline{n}%
(x)\mu \delta E_{x}(t)+  \nonumber \\
&&qD\frac{d\delta n_{x}(t)}{dx}+\frac{\delta I_{x}(t)}{A}\text{,}
\label{FlucDriftDif}
\end{eqnarray}
where $\delta E_{x}(t)$ and $\delta n_{x}(t)$ refer to the fluctuations of
electric field and number density at point $x$, respectively, and $\delta
I(t)$ refers to the fluctuations of the total current. In Eq.(\ref
{FlucDriftDif}), consistently with the low-frequency limit taken here, the
displacement current is neglected. Moreover, $\delta I_{x}(t)$ is a Langevin
source associated with the fluctuations of the current induced by the
diffusion of carriers inside the sample. It has zero mean and low frequency
spectral density,\cite{Ziel74,Vliet94,Bonani01} 
\begin{equation}
2\int_{-\infty }^{+\infty }dt\left\langle \delta I_{x}(t)\delta
I_{x^{^{\prime }}}(t^{\prime })\right\rangle =K(x)\delta (x-x^{\prime })%
\text{,}  \label{Correlation}
\end{equation}
with brackets indicating ensemble average, and 
\begin{equation}
K(x)=4qAk_{B}T\mu \overline{n}(x)=4Aq^{2}D\overline{n}(x),  \label{K(x)}
\end{equation}
where in the last equality use is made of the Einstein relation. Finally,
the linearized Poisson equation is given by 
\begin{equation}
\frac{d\delta E_{x}(t)}{dx}=-\frac{q}{\epsilon }\delta n_{x}(t)\text{,}
\label{FlucPoisson}
\end{equation}
and the linearized boundary conditions by (see appendix \ref
{metal-semiconductor}) 
\begin{equation}
\delta n_{L}(t)=\delta n_{0}(t)=0.  \label{FlucBc}
\end{equation}

In analogy with the case of the transport equations, we can combine Eqs.(\ref
{FlucDriftDif}) and (\ref{FlucPoisson}) into a single equation for the
electric field fluctuation of the form 
\begin{eqnarray}
\frac{\delta I(t)-\delta I_{x}(t)}{A} &=&-D\frac{d^{2}\delta E_{x}(t)}{dx^{2}%
}-\mu \overline{E}(x)\frac{d\delta E_{x}(t)}{dx}  \nonumber \\
&&-\mu \delta E_{x}(t)\left[ \frac{d\overline{E}(x)}{dx}-\frac{qN_{D}}{%
\epsilon }\right],  \label{FlucdE(x)}
\end{eqnarray}
with boundary conditions 
\begin{equation}
\left. \frac{d\delta E_{x}(t)}{dx}\right| _{0}=\left. \frac{d\delta E_{x}(t)%
}{dx}\right| _{L}=0.  \label{FlucBcE_x}
\end{equation}

\section{Stationary spatial profiles and current-voltage characteristics}

\label{Solution} In this section we present the results concerning the
transport properties for the structure under study calculated from the model
presented in Sec. \ref{Model_Transport}. In particular we will focus on the
transition from homogeneous to inhomogeneous conditions. To this end, for
given properties of the semiconductor we will vary the contact density from $%
n_{c}=N_{D}$ (corresponding to an homogeneous condition with Ohmic contacts)
to values $n_{c}>N_{D}$ (corresponding to inhomogeneous conditions with
injecting contacts). Note, that for given properties of the semiconductor,
to vary $n_{c}$ corresponds to vary the contact barrier height $\phi _{bn}$,
i.e. to change the material for the metal contact.

In what follows, the results are conveniently discussed in terms of the
values of two dimensionless parameters 
\begin{equation}
l=L/L_{D}\text{ and }\alpha =n_{c}/N_{D}\text{,}
\end{equation}
with $L_{D}=\left( k_{B}T\epsilon /q^{2}N_{D}\right) ^{1/2}$ being the Debye
screening length associated with the semiconducting material. The parameter $%
l$ depends only on the properties of the given semiconductor, while $\alpha $
depends also on the contact properties, thus being independent variables.
Since the non-equilibrium current noise properties of devices with
homogeneous stationary free carrier density profiles were shown to change
qualitatively when passing from $l<1$ to $l>1$, in what follows we will
treat these two cases separately.

\subsection{Homogeneous solution}

\label{Trans_homo} For $\alpha =1$ ($n_{c}=N_{D}$), and any value of $l$,
there exists a trivial homogeneous solution to the model presented in Sec.%
\ref{Model_Transport}, namely 
\begin{equation}
\overline{n}(x)=N_{D},\quad \overline{E}(x)=\frac{\overline{I}}{qA\mu N_{D}}.
\end{equation}
Accordingly, the current-voltage ($I-V$) characteristics is linear and
satisfies Ohm's law $\overline{I}=V/R_{bulk}$, with $R_{bulk}=L/(qA\mu
N_{D}) $ being the semiconductor resistance.

\subsection{Inhomogeneous solution}

\label{Trans_inhomo} For $\alpha >1$ ($n_{c}>N_{D}$) the solution to the
model presented in Sec. \ref{Model_Transport} is spatially inhomogeneous and
can not be obtained in a closed analytical form. Only under asymptotic
conditions it is possible to derive approximate analytical expressions\cite
{Lampert72,Ziel79} (see also appendix \ref{Asymptotic}). In any case, an
exact solution of the transport model can be obtained by numerical
integration following this procedure. One first solves numerically Eq.(\ref
{DDE_x}) subject to the boundary conditions in Eq.(\ref{BcE_x}). Having
found the electric field profile $\overline{E}(x) $, the carrier density
profile $\overline{n}(x)$ is obtained from the Poisson equation (\ref
{Poisson}). Furthermore, for a given value of the current $\overline{I}$,
the applied voltage is found as $\overline{V}=\int_{0}^{L}dx\overline{E}(x)$%
, from where the $I-V$ characteristic is obtained.

In what follow we discuss the main features of the inhomogeneous solution.
To this end we will consider two particular values of $l$, namely $l=0.1$
and $l=50$, as representative examples of the behavior observed for $l<1$
and $l>1$, respectively. In both cases, the values of $\alpha $ will be
varied in the range $1 \leq \alpha \leq 10^{4}$, thus allowing us to explore
both slightly and highly inhomogeneous situations.

\begin{figure}[tbp]
\centerline{\epsfxsize=8cm \epsffile{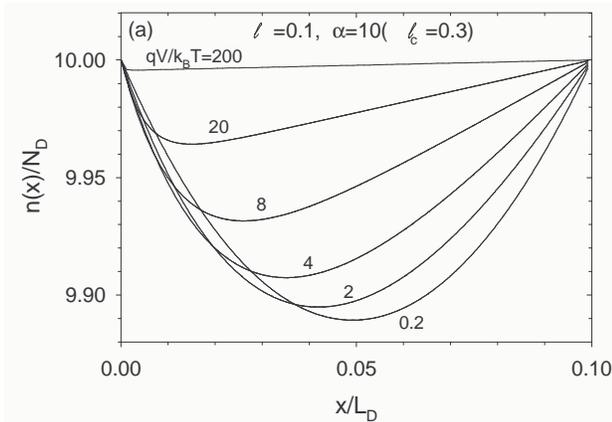}}
\centerline{\epsfxsize=8cm \epsffile{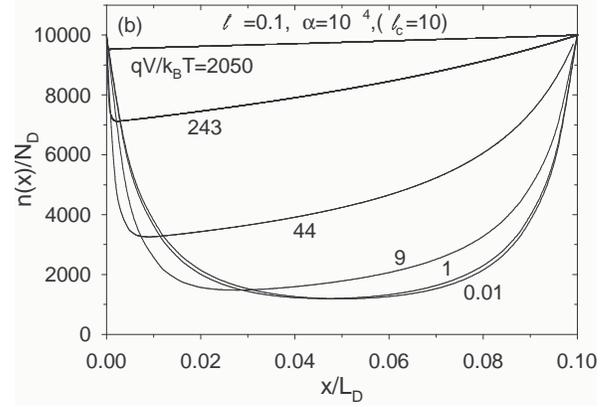}}
\caption{Free carrier density profiles normalized to the donor density for
several values of the applied bias, for $l=0.1$ and (a) $\alpha=10$ ($%
l_c=0.3 $) and (b) $\alpha=10^4$ ($l_c=10$).}
\label{FigProfiles01}
\end{figure}

Figure \ref{FigProfiles01} displays the stationary free-carrier density
profiles for $l=0.1$ for several values of the applied bias and for $%
\alpha=10$ (Fig. \ref{FigProfiles01}a) and $\alpha =10^{4}$ (Fig. \ref
{FigProfiles01}b). As can be seen, the stationary free carrier density
profiles are inhomogeneous and depend on the applied bias value.

For $\alpha =10$ (Fig.\ref{FigProfiles01}a) the degree of inhomogeneity is
rather small being at most of about $1\%$, and the free carrier density
departs more significantly from $n_{c}$ the lower the value of the applied
voltage. The physical reason for this behavior can be found in the fact that
the characteristic Debye screening length of the sample, roughly determined
by $n_{c}$, i.e. $L_{Dc}=\left( k_{B}T\epsilon /q^{2}n_{c}\right) ^{1/2}$,
is larger than the sample length $L$ ($l_{c}\equiv L/L_{Dc}=l\alpha
^{1/2}=0.1\times 10^{1/2}\sim 0.3$) thus not allowing the free carrier
density to relax from its contact value $n_{c}$ to its bulk value $N_{D}$,
i.e. to the local charge neutral state. Similar behaviors appears for other
values of $l<1$ and $\alpha >1$ as long as $l_{c}=l\alpha ^{1/2}\lesssim 1$.

For $\alpha =10^{4}$ (Fig.\ref{FigProfiles01}b) the degree of inhomogeneity
becomes appreciable being up to about a factor of $10$. For low applied bias 
$\overline{n}(x)$ is almost symmetric and departs significantly from $n_{c}$%
. The reason is that the characteristic Debye screening length, which can be
roughly approximated by $L_{Dc}$, is shorter than the sample length $L$ ($%
l_{c}=l\alpha ^{1/2}=0.1\times 10^{2}=10$), thus allowing relaxation towards
the charge neutral state. However, due to the small value of $l_{c}$,
complete relaxation of $\overline{n}(x)$ to the value $N_{D}$ is not
reached. At increasing applied bias, a net injection of carriers takes place
resulting in an increase of the values of $\overline{n}(x)$ and in a rather
asymmetric profile. Finally, for the highest applied biases carrier
injection no longer takes place, and the density distribution is almost
homogeneous with a value close to the contact value $n_{c}$. Similar
behaviors are observed for other values of $l<1$ and $\alpha >1$ as long as $%
l_{c}=l\alpha ^{1/2}>1$.

\begin{figure}[tbp]
\centerline{\epsfxsize=8cm \epsffile{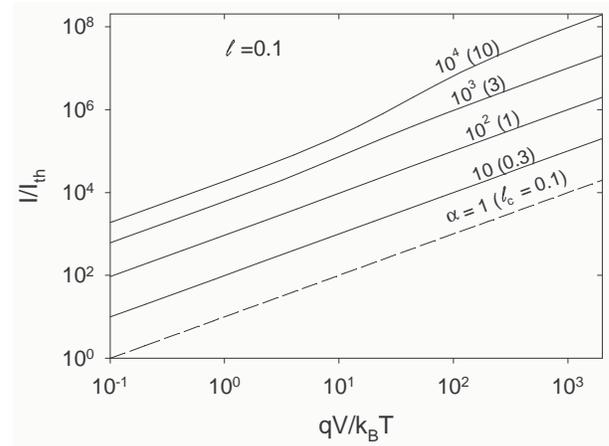}}
\caption{Current voltage characteristics for $l=0.1$ and several values of $%
\alpha$ (the corresponding value of $l_c$ is displayed for each curve). The
dashed line corresponds to the homogeneous solution.}
\label{FigIV01}
\end{figure}

The current-voltage characteristics for $l=0.1$ and for $\alpha $ ranging
from $1$ to $10^{4}$ are displayed in Fig. \ref{FigIV01}. For $\alpha \leq
10^{2}$, the $I-V$ characteristics is found to remain linear with a
resistance given by $R_{c}=L/(qA\mu n_{c})$. This linear behavior can be
understood by noting that for these values of $\alpha $ the free carrier
density profile is quasi-homogeneous with a value approximately equal to $%
n_{c}$ and almost independent of the applied bias (see Fig. \ref
{FigProfiles01}a). Similar behaviors are found for other values of $l<1$ and 
$\alpha >1$ as long as $l_{c}=l\alpha ^{1/2}\lesssim 1$. For $\alpha =10^{3}$
and $10^{4}$ the $I-V$ characteristics in Fig. \ref{FigIV01} displays a
super-linear behavior in an intermediate range of applied voltages, while it
is linear at the lowest and highest applied bias. The resistance at the
lowest bias, $R_{low}$, satisfies $R_{c}\leq R_{low}\leq $ $R_{bulk}$, since
in this range of bias the free carrier density profile is almost independent
of bias and satisfies $n_{c}\leq \overline{n}(x)\leq N_{D}$ (see Fig.\ref
{FigProfiles01}b). Furthermore, the resistance for high applied bias $%
R_{high}$ is almost equal to $R_{c}$, since in this range of bias the
density profiles is almost independent of bias and given by $\overline{n}%
(x)\sim n_{c}$ (see Fig. \ref{FigProfiles01}b). Finally, the superlinear
behavior is due to the net injection of carriers taking place in this bias
regime (see Fig. \ref{FigProfiles01}b) thus resulting in an increase of the
conductivity of the sample. Similar behaviors appear for other values of $l<1
$ and $\alpha >1$ as long as $l_{c}=l\alpha ^{1/2}>1$.

\begin{figure}[tbp]
\centerline{\epsfxsize=8cm \epsffile{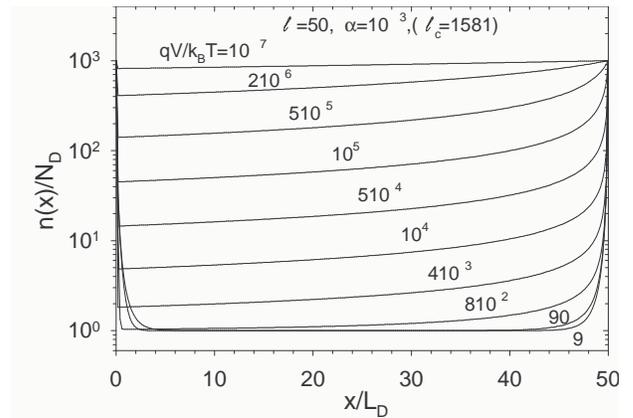}}
\caption{Free carrier density profiles normalized to the donor density for $%
l=50$ and $\alpha=10^3$ ($l_c=1581$), for several values of the applied
bias. }
\label{FigProfiles50}
\end{figure}

Let us now consider the case of $l=50$. Figure \ref{FigProfiles50} reports
the stationary free carrier density profile for $\alpha =10^{3}$ and several
values of the applied bias. The qualitative behavior of the profiles is
similar to those observed in Fig. \ref{FigProfiles01}b, since both cases
correspond to situations in which $l_{c}>1$ (in the present case $%
l_{c}=l\alpha ^{1/2}=50\times 10^{3/2}\sim 1581$). The main difference
between these figures is that in Fig. \ref{FigProfiles50} relaxation to the
local charge neutral state can take place for low bias, since the value of $%
l_{c}$ is high enough. Similar behaviors are observed for other values of $%
l>1$ and $\alpha >1$.

\begin{figure}[tbp]
\centerline{\epsfxsize=8cm \epsffile{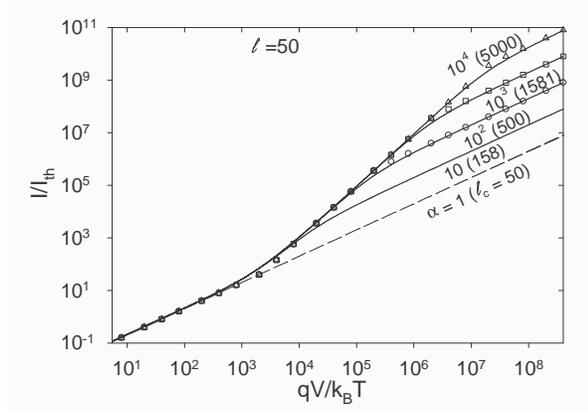}}
\caption{Current voltage characteristics for $l=50$ and several values of $%
\alpha$ (the corresponding value of $l_c$ is displayed for each curve). The
dashed line corresponds to the homogeneous solution. The symbols correspond
to the asymptotic expression of the $I-V$ characteristics given in Eq. (\ref
{IV_asymp}).}
\label{FigIV50}
\end{figure}

The current-voltage characteristics for $l=50$ and several values of $\alpha 
$ are plotted in Fig. \ref{FigIV50}. The $I-V$ characteristics is linear at
the lowest bias, super-linear at intermediate bias and again linear at the
highest bias. This behavior is similar to that reported in Fig. \ref{FigIV01}
for the curves with $\alpha =10^{3}$ and $10^{4}$, since both cases
correspond to situations where $l_{c}>1$.

In the present case ($l>1$) one can reach an asymptotic limit for $l\gg 1$
and $\alpha \gg 1$ in which the following analytical asymptotic expression
for the $I-V$ curve can be derived,  (see appendix \ref{Asymptotic})
\begin{equation}
\frac{\overline{I}}{I_{th}}\sim \left\{ 
\begin{array}{l}
\frac{1}{l}\frac{qV}{k_{B}T}\text{,\qquad \qquad }0\leq \frac{qV}{k_{B}T}%
\lesssim l^{2} \\ 
\frac{9}{8}\frac{1}{l^{3}}\left( \frac{qV}{k_{B}T}\right) ^{2}\text{, \quad }%
l^{2}\lesssim \frac{qV}{k_{B}T}\lesssim \alpha l^{2} \\ 
\frac{\alpha }{l}\frac{qV}{k_{B}T}\text{,\qquad \qquad }\alpha l^{2}\lesssim 
\frac{qV}{k_{B}T}
\end{array}
\right.   \label{IV_asymp}
\end{equation}
where we have defined $I_{th}=lk_{B}T/(qR_{bulk})$. Equation (\ref{IV_asymp}%
) gives a good description of the $I-V$ curves for $l\gtrsim 10$ and $\alpha
\gtrsim 100$, as can be seen in Fig. \ref{FigIV50} where the symbols
represent Eq. (\ref{IV_asymp}).\cite{lowalpha} Note that in the asymptotic
limit, the low bias resistance is given by $R_{low}=$ $R_{bulk}$,
independent from any contact parameter since at low bias a state of local
charge neutrality is achieved in the bulk of the sample (see Fig. \ref
{FigProfiles50}). Furthermore, in the intermediate bias regime, the $I-V$
characteristics display a quadratic bias dependence, in agreement with the
well-known Mott and Gurney law for space charge transport in diffusive
conductors.\cite{Lampert72} In this bias range, a strong injection of
carriers from the contacts takes place, as illustrated in Fig. \ref
{FigProfiles50}. Finally, at the highest voltages the resistance is given by 
$R_{high}=$ $R_{c}$, since the density profile is almost homogeneous and
equal to $n_{c}$ (see Fig.\ref{FigProfiles50}).

\section{Current noise properties}

\label{Results_Noise} We characterize the low frequency current fluctuation
properties by means of the low frequency current spectral density defined as 
\begin{equation}
S_{I}(0)=2\int_{-\infty }^{+\infty }\left\langle {\delta I(0)\delta I(t)}%
\right\rangle dt.  \label{SIdef}
\end{equation}
For the model presented in Sec. \ref{Model_Fluctuations}, $S_{I}(0)$ can be
given a closed analytical expression that takes into account in an exact way
the effects of the diffusion current on the non-equilibrium current
fluctuations and hence is applicable to the whole range of system
parameters. Below we discuss separately the cases corresponding to
homogeneous and inhomogeneous stationary density profiles, since they
involve rather different mathematics.

\subsection{Homogeneous stationary profiles}

\label{HomoNoise}

When the stationary profiles are homogeneous, Eq.(\ref{FlucdE(x)}) consists
of a second order differential equation with constant coefficients. The
solution of this equation, as well as an analytical expression for $S_{I}(0)$%
, has been derived recently:\cite{Gomila00b} 
\begin{eqnarray}
S_{I}(0) &=&\frac{4k_{B}T}{R_{bulk}}+  \nonumber \\
&&K\frac{(\lambda _{2}^{2}-\lambda _{1}^{2})}{2L^{2}\lambda _{1}^{2}\lambda
_{2}^{2}}\frac{\left( e^{\lambda _{1}L}-1\right) \left( e^{\lambda
_{2}L}-1\right) }{\left( e^{\lambda _{2}L}-e^{\lambda _{1}L}\right) ^{2}}%
\times  \label{Sihomogen} \\
&&\left[ \lambda _{2}\left( e^{\lambda _{2}L}+1\right) \left( e^{\lambda
_{1}L}-1\right) -\right.  \nonumber \\
&&\left. \lambda _{1}\left( e^{\lambda _{1}L}+1\right) \left( e^{\lambda
_{2}L}-1\right) \right]  \nonumber
\end{eqnarray}
where $K=4qAk_{B}T\mu \overline{n}$, and 
\begin{equation}
\lambda _{1,2}=-\frac{1}{2L_{E}}\left( 1\pm \sqrt{1+4\frac{L_{E}^{2}}{%
L_{D}^{2}}}\right) \text{;}  \label{lambda}
\end{equation}
with $L_{E}=k_{B}TL/(qV)$. In Eq.(\ref{lambda}) the subscript $1$
corresponds to the plus term and the subscript $2$ to the minus term.

For $l<1$, Eq. (\ref{Sihomogen}) takes the form \cite{Gomila00b}

\begin{equation}
S_{I}(0)=2q\overline{I}\coth (qV/2k_{B}T).  \label{SIcoth}
\end{equation}
Equation (\ref{SIcoth}) predicts a standard cross-over between Nyquist noise
(Eq.(\ref{Nyquisteq})) and shot noise (Eq.(\ref{shotI})) for $qV/k_{B}T\sim
3 $. The physical reason for the appearance of shot noise is the absence of
long range Coulomb correlations due to the fact that the sample length $L$
is smaller than the Debye screening length $L_{D}$.\cite{Gomila00b}

For $l>1$ the noise properties change considerably. Accordingly, in the
limit $l\gg 1$ Eq. (\ref{Sihomogen}) is well approximated by\cite{Gomila00b} 
\begin{equation}
S_{I}(0)=S_{I}^{th}\left\{ 
\begin{array}{c}
\frac{4}{l}\text{ for }0\leq \frac{\overline{I}}{I_{th}}\lesssim l^{1/3} \\ 
\frac{2}{l^{2}}\left( \frac{\overline{I}}{I_{th}}\right) ^{3}\text{ for }%
l^{1/3}\lesssim \frac{\overline{I}}{I_{th}}\lesssim l \\ 
2\left( \frac{\overline{I}}{I_{th}}\right) \text{ for }l\lesssim \frac{%
\overline{I}}{I_{th}}
\end{array}
\right.   \label{SIhomo}
\end{equation}
where $S_{I}^{th}=qI_{th}$. The above expression gives: Nyquist noise at low
bias ($0\leq qV/k_{B}T\lesssim l^{4/3}$), a cubic dependence on current of
noise at intermediate bias ($l^{4/3}\lesssim qV/k_{B}T\lesssim l^{2}$) and
shot noise at the highest bias ($qV/k_{B}T\geq l^{2}$). The physical reason
for the appearance of shot noise in this case is the vanishing of long range
Coulomb correlations because of a drift transit time shorter than both the
diffusion transit time and the dielectric relaxation time.\cite{Gomila00b}

\subsection{Inhomogeneous stationary profiles}

When the stationary profile is inhomogeneous, Eq. (\ref{FlucdE(x)}) consists
of a second-order stochastic differential equation with non-constant
coefficients. To obtain an analytical solution of this equation, we use a
method developed in Refs. [\onlinecite{Bulashenko97,Gomila98,Gomila99}]. The
method is based on the fact that 
\begin{equation}
\rho (x)=d\overline{E}(x)/dx,  \label{rho}
\end{equation}
constitutes a particular solution of Eq. (\ref{FlucdE(x)}). On this basis
one can find an analytical expression for the electric field fluctuations $%
\delta E_{x}(t)$ that solves Eq. (\ref{FlucdE(x)}) and satisfies the
boundary condition in Eq. (\ref{FlucBcE_x}).\cite{Gomila99} From the
expression of the electric field fluctuations we evaluate the voltage
fluctuations, $\delta V(t)=\int_{0}^{L}\delta E_{x}(t)dx$. Following a
procedure similar to that outlined in Ref. [\onlinecite{Gomila99}], the
voltage fluctuations can be expressed as: 
\begin{equation}
\delta V(t)=\int_{0}^{L}\nabla Z(x)\delta I_{x}(t)dx,  \label{dV_t}
\end{equation}
with 
\begin{eqnarray}
\nabla Z(x) &=&\frac{\rho (x)}{\epsilon AD}\left[ \int_{x}^{L}\frac{E(\xi
)-E_{\Delta }}{\rho ^{2}(\xi )}e^{-\frac{q(\phi (x)-\phi (\xi ))}{k_{B}T}%
}d\xi \right.   \nonumber \\
&&\qquad \qquad +\left. \frac{(E_{L}-E_{\Delta })}{\rho _{L}\rho
_{L}^{\prime }}e^{-\frac{q(V+\phi (x))}{k_{B}T}}\right] ,  \label{DZx}
\end{eqnarray}
and 
\begin{equation}
E_{\Delta }=\frac{\int_{0}^{L}\frac{E(x)}{\rho ^{2}(x)}e^{\frac{q\phi (x)}{%
k_{B}T}}dx+\frac{E_{L}}{\rho _{L}\rho _{L}^{\prime }}e^{-\frac{qV}{k_{B}T}}-%
\frac{E_{0}}{\rho _{0}\rho _{0}^{\prime }}}{\int_{0}^{L}\frac{1}{\rho ^{2}(x)%
}e^{\frac{q\phi (x)}{k_{B}T}}dx+\frac{1}{\rho _{L}\rho _{L}^{\prime }}e^{-%
\frac{qV}{k_{B}T}}-\frac{1}{\rho _{0}\rho _{0}^{\prime }}},  \label{E_D}
\end{equation}
Here, $\phi (x)$ is the electric potential. The function $\nabla Z(x)$ is
referred to as the impedance field,\cite{Vliet75,shokley66,shiktorov01}
since it satisfies 
\begin{equation}
Z(0)=\int_{0}^{L}\nabla Z(x)dx=\left( \frac{d\overline{I}}{dV}\right) ^{-1}.
\label{Z2}
\end{equation}
From Eq. (\ref{dV_t}), the low frequency spectral density of the voltage
fluctuations is given by 
\begin{equation}
S_{V}(0)=4Aq^{2}D\int_{0}^{L}[\nabla Z(x)]^{2}\overline{n}(x)dx,  \label{SV}
\end{equation}
where use is made of Eq. (\ref{Correlation}) and (\ref{K(x)}). The low
frequency spectral density of current fluctuations is then evaluated from%
\cite{Vliet94} 
\begin{equation}
S_{I}(0)=\frac{S_{V}(0)}{Z(0)^{2}},  \label{SISV}
\end{equation}

Equations (\ref{DZx})-(\ref{SISV}) constitute an exact solution of the noise
model presented in Sec. \ref{Model} for the general case of an inhomogeneous
stationary profile. They represent the main result of this paper. Note, that
the solution incorporates the effects of the diffusion current without any
approximation, thus allowing us to investigate the transition from
homogeneous to highly inhomogeneous stationary profiles.

In order to evaluate Eqs. (\ref{DZx})-(\ref{SISV}) we will use the
steady-state profiles calculated in Sec. \ref{Trans_inhomo}. As for the case
of transport we will consider the values $l=0.1$ and $l=50$, as
representative examples of the behavior observed for $l<1$ and $l>1$.

\begin{figure}[tbp]
\centerline{\epsfxsize=8cm \epsffile{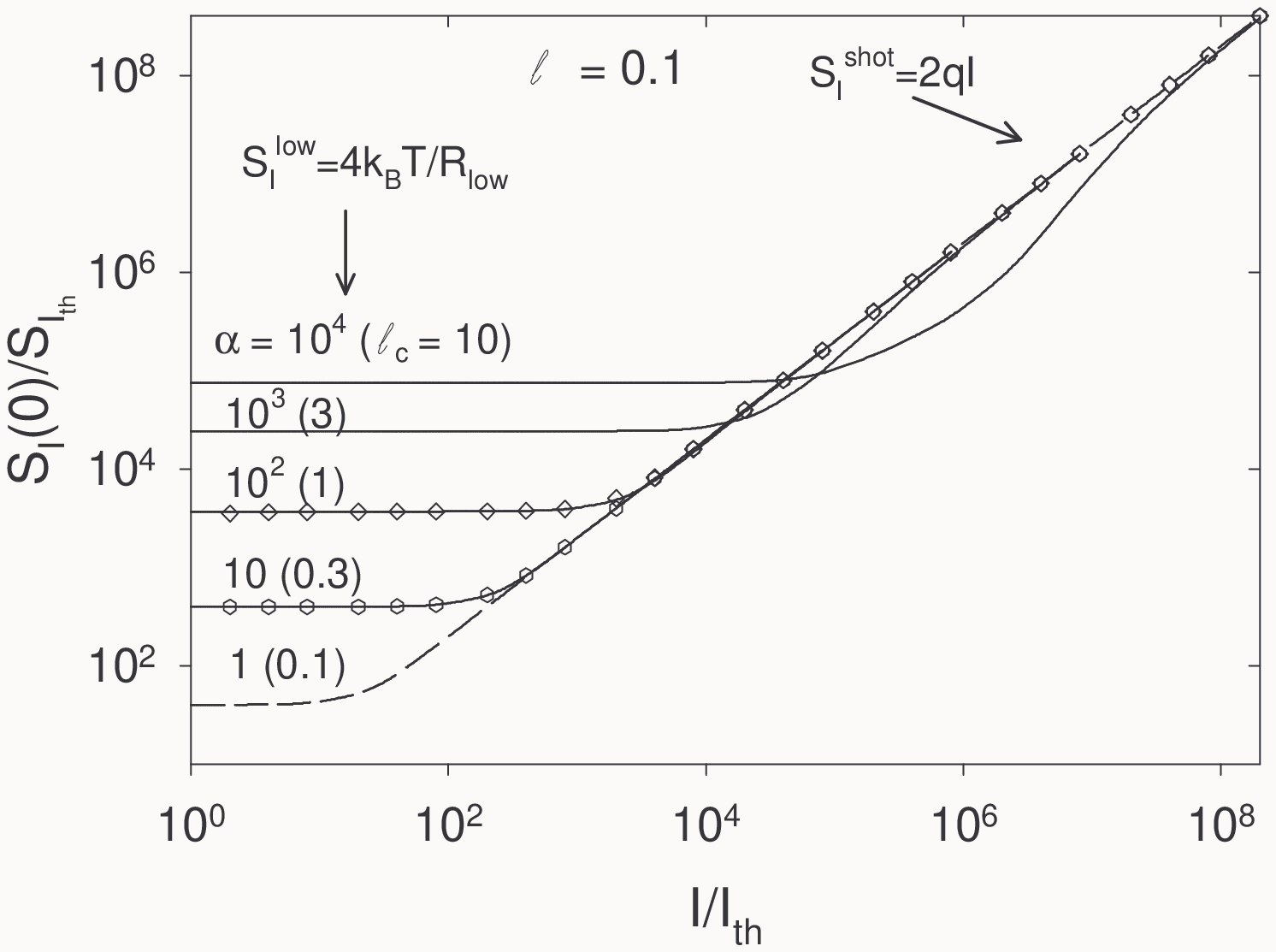}}
\caption{Low frequency current spectral density normalized to ${S_I}%
_{th}=qI_{th}$ as a function of the current normalized to $I_{th}$ for $%
l=0.1 $ and several values of $\alpha$ (the corresponding value of $l_c$ is
displayed for each curve). The dashed line corresponds to the homogeneous
solution. The symbols correspond to the coth-like expression in Eq. (\ref
{SIcoth}).}
\label{FigSI01}
\end{figure}

Figure \ref{FigSI01} reports the low frequency spectral density of current
fluctuations $S_{I}(0)$ as a function of the current as obtained from Eqs. (%
\ref{DZx})-(\ref{SISV}) for $l=0.1$ and several values of $\alpha $. For
comparison the results for the homogeneous case, as calculated from Eq. (\ref
{Sihomogen}), are also displayed (dashed line). We note that for the lowest
degree of inhomogeneity ($\alpha \leq 10^{2}$) the non-equilibrium current
noise results can be well approximated by Eq. (\ref{SIcoth}) (symbols in
Fig. \ref{FigSI01}). The reason is that for these values of $\alpha $ the
profiles are quasi-homogeneous with free carrier density $n_{c}$ and with
Debye screening length satisfying $l_{c}<1$, which are the conditions for
the validity of Eq.(\ref{SIcoth}). Similar behaviors are observed for other
values of $l<1$ and $\alpha >1$ for which $l_{c}<1$.

By contrast, for a high degree of inhomogeneity ($\alpha =10^{3}$ and $%
10^{4} $) the results deviate from Eq.(\ref{SIcoth}), and regions where the
current noise is suppressed below the shot noise limit are observed at
intermediate bias values. The bias values where shot noise suppression takes
place are found to correspond to those applied bias for which significant
net carrier injection from the contacts takes place (see Fig. \ref
{FigProfiles01}b). The physical origin of the shot noise suppression should
be traced back to the correlations induced by the long range Coulomb
interaction which are active in this range of bias. Similar behaviors are
observed for other values of $l<1$ and $\alpha >1$ for which $l_{c}>1$.

\begin{figure}[tbp]
\centerline{\epsfxsize=8cm \epsffile{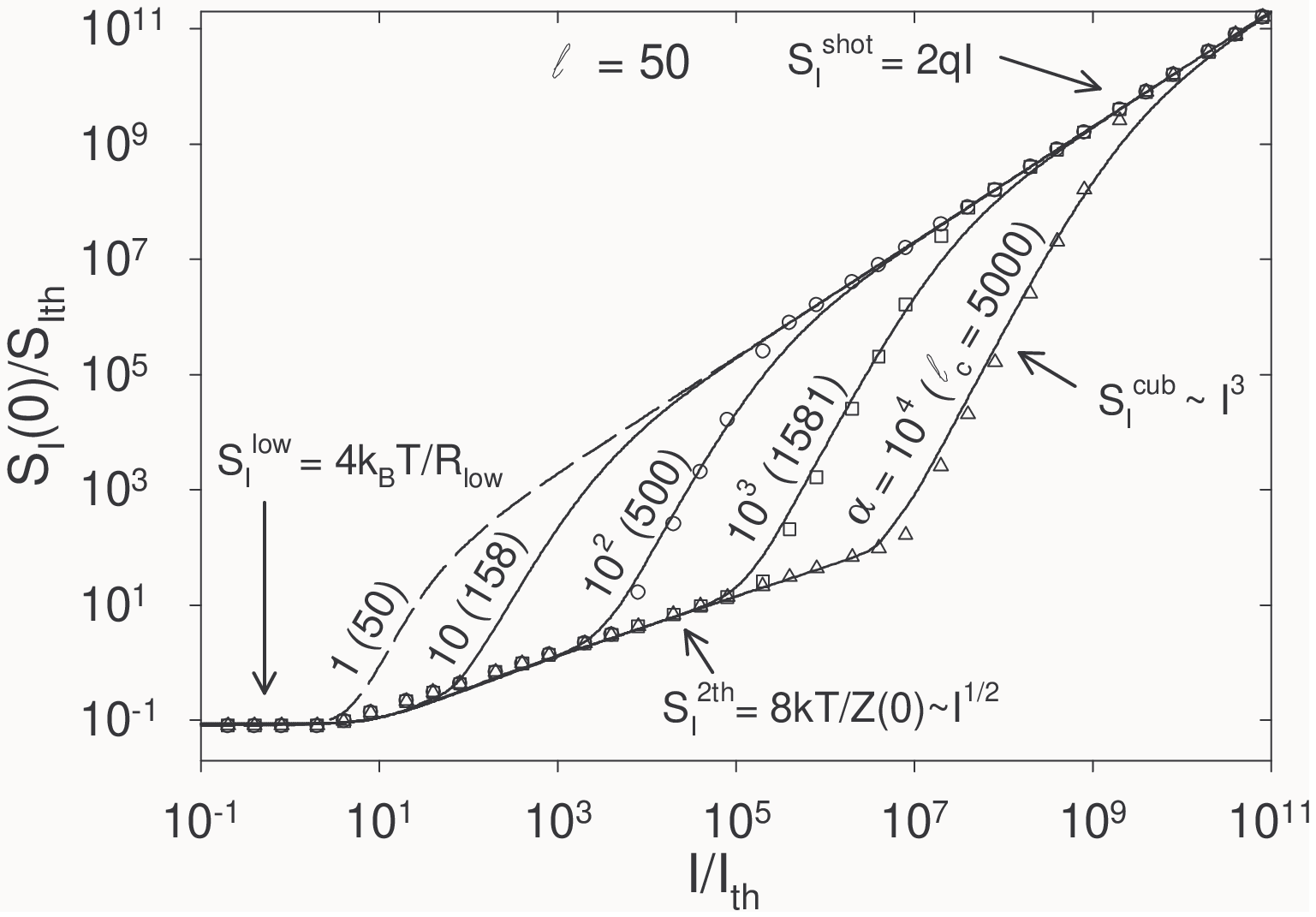}}
\caption{Low frequency current spectral density normalized to ${S_I}%
_{th}=qI_{th}$ as a function of the current normalized to $I_{th}$ for $l=50$
and several values of $\alpha$ (the corresponding value of $l_c$ is
displayed for each curve). The dashed line corresponds to the homogeneous
solution. The symbols correspond to the approximate expression in Eq. (\ref
{SI_asymp}).}
\label{FigSI50}
\end{figure}

Figure \ref{FigSI50} reports the low frequency spectral density of current
fluctuations $S_{I}(0)$ as a function of the current as obtained from Eqs. (%
\ref{DZx})-(\ref{SISV}) when $l=50$ and for several values of $\alpha $. For
comparison, the results for the homogeneous case, as calculated from Eq.(\ref
{Sihomogen}), are also displayed (dashed line). As can be seen, a
qualitative new behavior at intermediate current values is identified with
respect to the homogeneous behavior for all values of $\alpha $. The new
behavior is found to tend to a square root dependence on the current and to
cover broader current intervals as $\alpha $ is increased. This fact can be
quantitatively evaluated by means of the following explicit expression valid
for $l\gtrsim 10$ and $\alpha \gtrsim 100$ 
\begin{equation}
\frac{S_{I}(0)}{S_{I}^{th}}\sim \left\{ 
\begin{array}{l}
\frac{4}{l}\text{,\qquad \qquad \qquad }0\lesssim \frac{\overline{I}}{I_{th}}%
\lesssim \frac{l}{18} \\ 
\frac{12\sqrt{2}}{l^{3/2}}\left( \frac{\overline{I}}{I_{th}}\right) ^{1/2}%
\text{,}\,\,\,\,\frac{l}{18}\lesssim \frac{\overline{I}}{I_{th}}\lesssim
\left( \frac{9}{2}\alpha ^{8}l\right) ^{\frac{1}{5}} \\ 
\frac{8}{\alpha ^{4}l^{2}}\left( \frac{\overline{I}}{I_{th}}\right) ^{3}%
\text{,}\,\,\,\,\,\,\,\,\,\,\,\left( \frac{9}{2}\alpha ^{8}l\right) ^{\frac{1%
}{5}}\lesssim \frac{\overline{I}}{I_{th}}\lesssim \frac{\alpha ^{2}l}{2} \\ 
2\left( \frac{\overline{I}}{I_{th}}\right) \text{,}\qquad
\,\,\,\,\,\,\,\,\,\,\frac{\alpha ^{2}l}{2}\lesssim \frac{\overline{I}}{I_{th}%
}
\end{array}
\right.  \label{SI_asymp}
\end{equation}
(see Fig. \ref{FigSI50} where the symbols correspond to the approximate
expression). Otherwise, for $1<l<10$ and $1<\alpha <100$ the exact result
should be used.

\begin{figure}[tbp]
\centerline{\epsfxsize=8cm \epsffile{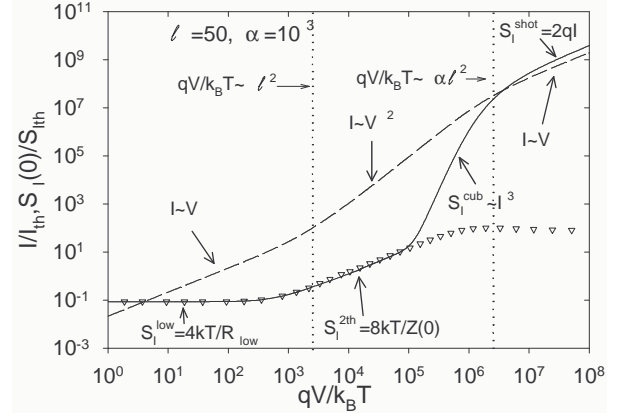}}
\caption{Electric current (dashed line) and low frequency current spectral
density (solid line) as a function of applied bias for $l=50$ and $%
\alpha=10^3$. The different behaviours are schematically indicated. The
current is normalized to $I_{th}$ and the spectral density to $S^{th}_I=q
I_{th}$. For comparison the results obtained from Eq. (\ref{Double}) are
also displayed as triangles.}
\label{FigSI_IV}
\end{figure}

The dependence of the noise power as the square root of the current
corresponds to the well known double thermal noise regime found in space
charge limited devices\cite{Vliet75,Nicolet75} (see Eq.(\ref{Isquare})). We
note that the double thermal noise behavior is restricted to current values
in the range $l/18\lesssim \overline{I}/I_{th}\lesssim \left( 4.5\alpha
^{8}l\right) ^{1/5}$. For higher currents the noise properties deviate from
the double thermal noise behavior and start resembling those corresponding
to homogeneous conditions (cubic current dependence followed by shot noise).
This fact is illustrated in Fig. \ref{FigSI_IV} where we compare the exact
results of the present paper (continuous line) with those of existing
theories for double thermal noise represented by Eq. (\ref{Double})
(triangles) for $l=50$ and $\alpha =10^{3}$. As can be seen, by neglecting
the diffusion current, existing theories can only be applied up to bias
values below the onset of the cubic region. At applied bias above this
onset, the diffusion current plays a relevant role and the exact theory
presented here must be used. It is worth remarking that for devices
operating under strong current injection conditions ($\alpha \gtrsim 10^{4}$
) the quasi-homogenous behavior predicted at the highest bias can be hardly
observed in practice because of hot-carrier effects.\cite{Reggiani85}
However, in the low current injection regime ($1<\alpha \lesssim 10^{4}$)
this behavior should be experimentally accessible as discussed in the
following suggested example.

\subsection{Example}

\label{Exemple}

As an example to illustrate the experimental accessibility of the
theoretical predictions described above we consider the following particular
case. As semiconductor material we consider high resistive $p$-type CdTe at
room temperature with a free carrier density (holes) $p=10^{8}\ $cm$^{-3}$.
These low values of the free carrier density can be obtained by means of
compensation.\cite{Straus97} The hole mobility is taken to be $\mu \sim 40$
cm$^{2}$/Vs, the effective hole mass $m^{*}=0.8\ m_{0}$, where $m_{0}$ is
the electron mass, and the dielectric constant $\epsilon =10.3\ \epsilon _{0}
$, with $\epsilon _{0}$ being the vacuum permittivity. The sample length is
assumed to be $L=4\ $mm and the cross sectional area $A=40$ mm$^{2}$. As
metal for the contacts we consider gold, which has been reported to form
almost ideal metal-semiconductor junctions on CdTe.\cite{Dharmadasa89} The
value for the Au/CdTe barrier height for holes, deduced from that for
electrons, is $\phi _{bp}=0.55\ V$, where use has been made of the
relationship $\phi _{bp}=E_{g}-\phi _{bn}$ with $E_{g}=1.48$ eV and $\phi
_{bn}=0.93$ V.\cite{Dharmadasa89} According to the value of $\phi _{bp}$ the
contact density for holes is $p_{c}=10^{10}\ $cm$^{-3}$, where we used that $%
p_{c}=N_{V}\exp [-q\phi _{bp}/(k_{B}T)]$, with $N_{V}=1.8$ $10^{19}$cm$^{-3}$
being the CdTe effective density of the states in the valence band at room
temperature. The metal contact described above forms an ohmic injecting
contact in the terms described in the present paper (see also at the end of
appendix \ref{metal-semiconductor}.)

\begin{figure}[tbp]
\centerline{\epsfxsize=8cm \epsffile{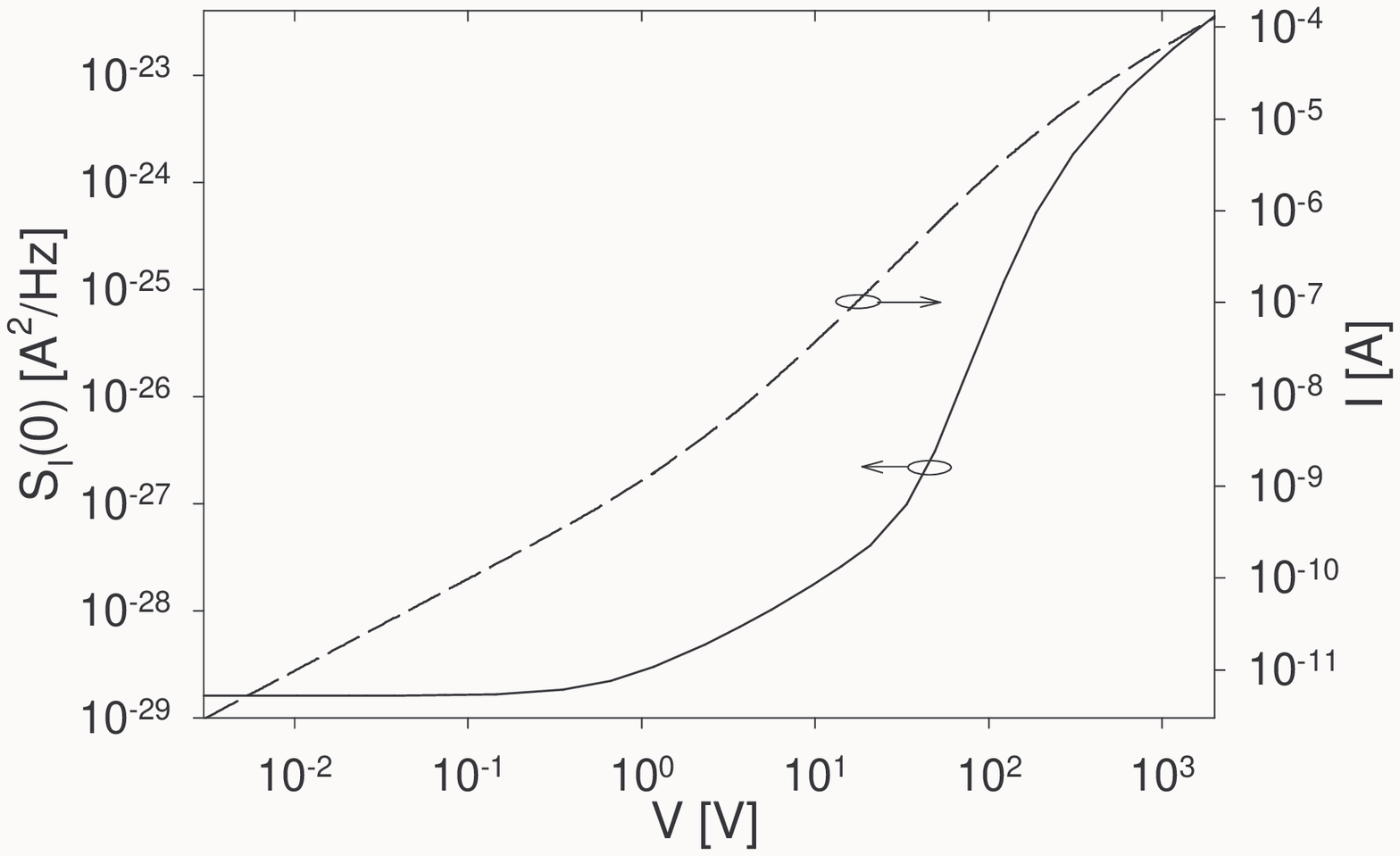}}
\caption{Electric current (dashed line) and low frequency current spectral
density (solid line) as a function of applied bias for a highly resistive $%
p- $type CdTe metal-semiconductor-metal structure. Parameters: free carrier
density $10^{8}$ cm$^{-3}$, sample length $4$ mm, cross sectional area $40$
mm$^{2}$, temperature $300$ K and contact density $10^{10}$ cm$^{-3}$.}
\label{FigCdTe}
\end{figure}

For the set of parameters considered above, we have $L_{D}=0.4$ mm and $%
L_{Dc}=0.04$ mm, from where $l=L/L_{D}=10$, $\alpha =n_{c}/N_{D}=100$ and $%
l_{c}=L/L_{Dc}=100$. Figure \ref{FigCdTe} displays the calculated current $I$
(right axis) and low frequency current spectral density (left axis) as a
function of the applied. According to the calculations, the predicted $I-V$
characteristic is linear up to bias voltages around $0.6$ V. For higher bias
it tends to be quadratic, up to $V\sim 300$ V, from where it returns to
linearity. Furthermore, the calculated current spectral density $S_{I}(0)$
displays the Nyquist thermal value up to voltages around $0.3$ V. Then, it
increases with voltage according to the double thermal noise behavior up to
around $20$ V. At further increasing voltages, $S_{I}(0)$ increases sharply
with voltage according to the cubic dependence with applied current up to $%
V\sim 1kV$, where shot noise appears. It is worth noting that the calculated
values of the low frequency current spectral density are well inside the
range of experimental accessibility (state of the art correlation spectrum
analyzers\cite{Sampietro99} are able to reach noise levels as low as $%
10^{-29}$ A$^{2}$/Hz). Moreover, the electric fields reach maximum values up
to $E_{av}=V/L=5$ kV/cm, which are still below that for the onset of hot
electron effects in $p$-type CdTe. We conclude that such an example shows
that the theoretical predictions presented in this paper are accessible to
an experimental confirmation.

\section{Conclusions}

\label{Conclusions}

We have carried out an analytical theory of transport and current
fluctuation properties in metal-semiconductor-metal structures made of
highly resistive semiconductors. The theory includes the effects of the
diffusion current in an exact way, thus allowing us to study the whole range
of physical conditions concerning the strength of the applied bias and the
level of carrier injection from the contacts. It is shown that in the low
carrier injection limit for which $l_{c}=L/L_{Dc}<1$, where $L$ is the
sample length and $L_{Dc}$ the Debye screening length associated with the
free carrier density at the metal-semiconductor interface $n_{c}$, the
structure behaves like a linear resistor with low frequency noise properties
given by $S_{I}(0)=2q\overline{I}\coth [qV/(2k_{B}T)]$. In the intermediate
carrier injection regime, roughly determined by $1<l_{c}\lesssim 100$, the
structure displays a linear-superlinear-linear current-voltage
characteristics. In this regime, the current spectral density displays a
cross-over from Nyquist noise to shot noise mediated by a region depending
first as the square root of the current (double thermal noise) and then as
the third power of the current. Finally, under strong carrier injection
conditions $l_{c}\gg 1$ the standard theory of space charge limited diodes
is recovered. In this limit the current-voltage characteristics is first
linear and then quadratic. Accordingly, the current spectral density
displays Nyquist thermal noise at low bias followed by double thermal noise
at higher applied bias. We suggest that high resistive p-type CdTe is one of
the best suited materials to provide an experimental test of the theoretical
predictions in the small and moderated injection regimes.

\section{Acknowledgments}

Partial support from the MCyT-Spain through the Ramon y Cajal program and
projects Nos. BFM2000-0624 and BFM2001-2159 and from the Italy-Spain Joint
Action of the MIUR-Italy (Ref. IT109) and MCyT-Spain (Ref. HI2000-0138) is
gratefully acknowledged. Prof. T. Gonzalez of Salamanca University is
acknowledged for the stimulating discussions carried out on the subject.

\appendix

\section{Metal-semiconductor model contact}

\label{metal-semiconductor} In this appendix we justify the boundary
conditions used along the paper for both the transport, Eq. (\ref{Bc}) and
the noise properties, Eq. (\ref{FlucBc}). Following Refs. [%
\onlinecite{Sze81,Gomila96,Gomila97}], the general boundary conditions for
the transport through a metal-semiconductor contact read 
\begin{equation}
\overline{n}(0)=\frac{\overline{I}+I_{s}^{0}}{qAv_{r}^{0}}\text{;\quad }%
\overline{n}(L)=\frac{-\overline{I}+I_{s}^{L}}{qAv_{r}^{L}},  \label{Bc_MS}
\end{equation}
with $v_{r}^{0}$ and $I_{s}^{0}$ (res. $v_{r}^{L}$ and $I_{s}^{L}$) being
the recombination velocity and saturation current, respectively, of the
contact located at $x=0$ (res. $x=L$). The currents $I_{s}^{L}$ and $%
I_{s}^{0}$ are given by $I_{s}^{0}=qAv_{r}^{0}n_{0}^{eq}$ and $%
I_{s}^{L}=qAv_{r}^{L}n_{L}^{eq}$, where 
\begin{equation}
n_{0}^{eq}=N_{C}e^{-\frac{q\phi _{bn}^{0}}{k_{B}T}}\text{;\quad }%
n_{L}^{eq}=N_{C}e^{-\frac{q\phi _{bn}^{L}}{k_{B}T}},  \label{n0eq}
\end{equation}
Here, $N_{C}$ is the effective density of states in the conduction band and $%
\phi _{bn}^{0}$ (res. $\phi _{bn}^{L}$) is the barrier height at contact $%
x=0 $ (res. $x=L$). For the case of ideal metal-semiconductor junctions, the
barrier heights are given by $\phi _{bn}^{0}=\phi _{m}^{0}-\chi $ and $\phi
_{bn}^{L}=\phi _{m}^{L}-\chi $, with $\chi $ being the semiconductor
affinity and $\phi _{m}^{0}$ (res. $\phi _{m}^{L}$) the work function of the
metal located at $x=0$ (res. $x=L$). For highly resistive materials the
diffusion approximation\cite{Sze81} can be applied. In this approximation it
is assumed that the current takes values much smaller that the thermionic
current, i.e. $\overline{I}\ll I_{s}$. This condition implies \cite{Gomila98}
$\overline{I}/I_{th}\ll \overline{I}_{s}/I_{th}=\alpha /\beta $, where 
\begin{equation}
\beta =\mu \sqrt{\frac{2\pi N_{D}m^{*}}{\epsilon }};\quad \alpha =\frac{%
n^{eq}}{N_{D}}.  \label{alfa}
\end{equation}
In this limit, one can approximate Eq.(\ref{Bc_MS}) by 
\begin{equation}
\overline{n}(0)\sim \frac{I_{s}^{0}}{qAv_{r}^{0}}=n_{0}^{eq}\text{;\quad }%
\overline{n}(L)\sim \frac{I_{s}^{L}}{qAv_{r}^{L}}=n_{L}^{eq},
\end{equation}
which for the case of a symmetric structure correspond to the boundary
conditions in Eq.(\ref{Bc}). This result also implies that the resistance of
the metal-semiconductor interface is negligible small in comparison to the
bulk resistance. As a result, the fluctuations generated at the
metal-semiconductor interface can be completely neglected. This is
equivalent to approximate\cite{Gomila98} 
\begin{equation}
\delta n_{0}(t)\sim 0\text{; \quad }\delta n_{L}(t)\sim 0,
\end{equation}
in agreement with the boundary conditions used along the paper Eq. (\ref
{FlucBc}).

For the concrete example considered in Sec.\ref{Exemple}, one has $\beta =9\
10^{-6}\ll 1$, which allows us to apply the diffusive approximation up to
currents satisfying the condition $I_{s}/I_{th}=\alpha /\beta =1.1\ 10^{7}$,
well above the current values required to observe the different behaviors
here identified.

\section{Asymptotic transport theory}

\label{Asymptotic} In this appendix we derive the asymptotic expression for
the $I-V$ characteristics used in Eq.(\ref{IV_asymp}) of Sec. \ref
{Trans_inhomo}. To this purpose, we start by considering the limit when $%
l\gg 1$. In this limit, one can neglect the diffusive contribution in Eq.(%
\ref{DDE_x}), thus obtaining a first order differential equation for the
electric field of the form 
\begin{equation}
-\overline{E}(x)\left[ \frac{d\overline{E}(x)}{dx}-1\right] =\overline{I},
\label{DDasy}
\end{equation}
subject to the boundary condition $E(L)=\overline{I}/\alpha $. In equation (%
\ref{DDasy}), and along this appendix, we use dimensionless variables to
simplify the expressions. Accordingly, we take $\overline{E}\rightarrow 
\overline{E}/E_{th}$, $\overline{I}\rightarrow \overline{I}/I_{th}$, $%
x\rightarrow x/L_{D}$, $L\rightarrow L/L_{D}$, $\overline{U}(x)\rightarrow 
\overline{U}(x)/k_{B}T$, where $E_{th}=\left( k_{B}TN_{D}/\epsilon \right)
^{1/2}$.Equation (\ref{DDasy}) can be easily integrate between $x=0$ and $x=L
$ to give the following relation 

\begin{equation}
\overline{E}_{L}+\overline{I}\ln \left( 1-\frac{\overline{E}_{L}}{\overline{I%
}}\right) -\overline{E}_{0}+\overline{I}\ln \left( 1-\frac{\overline{E}_{0}}{%
\overline{I}}\right) =L.  \label{E_L}
\end{equation}
where $\overline{E}_{L}=\overline{E}(L)$ and $\overline{E}_{0}=\overline{E}%
(0)$.  On the other hand, by using again Eq.(\ref{DDasy}), one can derive an
explicit expression for the potential energy difference $\overline{U}(x)-%
\overline{U}_{L}=\int_{L}^{x}\overline{E}(x)dx$, which once evaluated at $x=0
$ gives for the applied bias, $V=U_{L}-U_{0}$, 
\begin{eqnarray}
\overline{V} &=&\frac{\overline{E}_{L}^{2}}{2}+\overline{I}\overline{E}_{L}+%
\overline{I}^{2}\ln \left( 1-\frac{\overline{E}_{L}}{\overline{I}}\right)  
\nonumber \\
&&-\frac{\overline{E}_{0}^{2}}{2}-\overline{I}\overline{E}_{0}-\overline{I}%
^{2}\ln \left( 1-\frac{\overline{E}_{0}}{\overline{I}}\right) .
\label{V1_asym}
\end{eqnarray}
To end up with an explicit expression for the $I-V$ characteristics, we
impose the boundary condition $\overline{E}_{L}=\overline{I}/\alpha $ in
Eqs.(\ref{V1_asym}) and (\ref{E_L}), thus giving the $I-V$ characteristics
in parametric form 
\begin{eqnarray}
\overline{V} &=&\frac{\left[ \frac{1}{2\alpha ^{2}}+\frac{1}{\alpha }+\ln
\left( 1-\frac{1}{\alpha }\right) -\frac{u_{0}^{2}}{2}-u_{0}-\ln \left(
1-u_{0}\right) \right] }{L^{2}\left[ \frac{1}{\alpha }+\ln \left( 1-\frac{1}{%
\alpha }\right) -u_{0}-\ln \left( 1-u_{0}\right) \right] ^{2}},
\label{I_u02} \\
\overline{I} &=&\frac{L}{\left[ \frac{1}{\alpha }+\ln \left( 1-\frac{1}{%
\alpha }\right) -u_{0}-\ln \left( 1-u_{0}\right) \right] },  \label{V_u02}
\end{eqnarray}
where the parameter $u_{0}=\overline{E}_{0}/\overline{I}$ satisfies the
condition $1/\alpha <u_{0}<1$. By further expanding the previous expression
for $\alpha \gg 1$ one then arrives at Eq. (\ref{IV_asymp}).


\end{multicols}


\begin{references}
\bibitem{lowalpha}  For values of $l\gtrsim 10$ and $\alpha <100$ one can
use the parametric expression for the $I-V$ derived in appendix \ref
{Asymptotic}, Eqs. (\ref{I_u02}) and (\ref{V_u02}).
\end{references}
\end{document}